# Direct atomic layer deposition of ultra-thin $Al_2O_3$ and $HfO_2$ films on gold-supported monolayer $MoS_2$

E. Schilirò[1*], S. E. Panasci[1], A.M. Mio[1], G. Nicotra[1], S. Agnello[2,1,3], B. Pecz[4], G.Z. Radnoczi[4], I. Deretzis[1], A. La Magna[1], F. Roccaforte[1], R. Lo Nigro[1], F. Giannazzo[1]

[1] CNR-IMM, Strada VIII, 5 95121, Catania, Italy

[2] Department of Physics and Chemistry, University of Palermo, Via Archirafi 36, 90143 Palermo, Italy

[3] ATeN Center, Viale delle Scienze Ed. 18, 90128 Palermo, Italy

[4] Institute for Technical Physics and Materials Science, Centre for Energy Research, HAS, 1121 Konkoly-Thege 29-33, Budapest, Hungary

*e-mail: emanuela.schiliro@imm.cnr.it

**Abstract**

In this paper, the atomic layer deposition (ALD) of ultra-thin films (<4 nm) of $Al_2O_3$ and $HfO_2$ on gold-supported monolayer (1L) $MoS_2$ is investigated, providing an insight on the mechanisms ruling the nucleation in the early stages of the ALD process. A preliminary multiscale characterization of large area 1L-$MoS_2$ exfoliated on sputter-grown Au/Ni films demonstrated: (i) a tensile strain (from 0.1 to 0.3%) and p-type doping (from $1\times10^{12}$ to $4\times10^{12}$ cm$^{-2}$) distribution at micro-scale; (ii) an almost conformal $MoS_2$ membrane to the Au grains topography, with some locally detached regions, indicating the occurrence of strain variations at the nanoscale; (iii) atomic scale variability (from ~4.0 to ~4.5 Å) in the Mo-Au atomic distances was detected, depending on the local configuration of Au nanograins. *Ab initio* DFT calculations of a free-standing $MoS_2$ layer and a simplified $MoS_2$/Au(111) interface model showed a significant influence of the Au substrate on the $MoS_2$ energy band structure, whereas small differences were accounted for the adsorption of $H_2O$, TMA (co-reactant, and Al-precursor, respectively) molecules, and a slight improved adsorption was predicted for TDMAHf

(Hf-precursor). This suggests a crucial role of nanoscale morphological effects, such as the experimentally observed local curvature and strain of the $MoS_2$ membrane, in the enhanced physisorption of the precursors. Afterwards the nucleation and growth of $Al_2O_3$ an $HfO_2$ films onto 1L-$MoS_2$/Au was investigated in detail, by monitoring the surface coverage as a function of the number (N) of ALD cycles, with N from 10 to 120. At low N values, a slower growth rate of the initially formed nuclei was observed for $HfO_2$, probably due to the bulky nature of the TDMAHf precursor as compared to TMA. On the other hand, the formation of continuous films was obtained in both cases for N>80 ALD cycles, corresponding to ~3.6 nm $Al_2O_3$ and ~3.1 nm $HfO_2$. Current mapping on these ultra-thin films by conductive-AFM showed, for the same applied bias, a uniform insulating behavior of $Al_2O_3$ and the occurrence of few localized breakdown spots in the case of $HfO_2$, associated to a less compact films regions. Finally, an increase of the 1L-$MoS_2$ tensile strain was observed by Raman mapping after encapsulation with both high-κ films, accompanied by a reduction in the PL intensity, explained by the effects of strain and the higher effective dielectric constant of the surrounding environment.

1. **Introduction**

Molybdenum disulfide (2H-$MoS_2$) is currently the most investigated two dimensional (2D) material in the transition metal dichalcogenides (TMDs) class. Due to its peculiar tunable semiconducting behavior, characterized by a direct band-gap of 1.8 eV for monolayer (1L) and an indirect band-gap of 1.2 eV for few layers and bulk films[1,2], it is very promising for logic and switching electronic devices. In this respect, the integration of high-κ dielectrics, such as $Al_2O_3$ and $HfO_2$, on $MoS_2$ surface proved to be essential for obtaining high performance in terms of field-effect mobility (100-500 $cm^2$/V s), sub-threshold swing (~ 70 mV/dec), and on/off ratio (~ $10^8$) in $MoS_2$-based field effect transistors [3,4,5]. Currently, the atomic layer deposition (ALD) represents the technique of choice for the growth of high-κ dielectrics in microelectronics, thanks to its peculiar working principle (consisting of a series of self-limited reaction cycles between precursors) that guarantees a nanometric thickness control, conformity, and uniformity on large areas [6]. However, the direct ALD on $MoS_2$ and other 2D materials, is typically hindered by the lack of out-of-plane bonds on the surface of van der Waals materials, which should act as nucleation sites for ALD precursors [7,8]. To circumvent this issue, different approaches have been explored to enhance the ALD growth on the 2D-materials, including the pre-deposition of seed-layers and pre-functionalization processes are the most used methods [9,10,11,12]. Actually, such approaches allow to achieve a uniform high-k dielectrics coverage on the MoS$_2$ surface. On the other hand, they affect the interface quality and ultimately limit the minimum deposited dielectric thickness. Besides using seed-layers and pre-functionalization treatments, the ALD process on the 2D materials can be also improved by adjusting the deposition parameters, including temperature, reactivity of co-reactants (H$_2$O, O$_2$-plasma, O$_3$) and precursor flow [13, 15]. The temperature of the ALD process has been found to be determinant for the uniformity of growth on MoS$_2$. In particular, temperatures lower than 150°C improve the Al$_2$O$_3$ coverage, by limiting the desorption phenomena of the ALD-precursor, i.e. trimethylaluminium (TMA) [8]. On the other hand, low deposition temperatures result in a poor structural and electrical quality of the dielectric, as a consequence of a decreased reactivity of ALD precursors [14]. Also resorting to more reactive ALD co-reactants, such as O$_3$ instead of H$_2$O [15], or using O$_2$-plasma pre-treatments of the MoS$_2$ surface [16], can be useful to optimize the uniformity of dielectrics on MoS$_2$. Unfortunately, such treatments can be detrimental for MoS$_2$ due to defects and MoO$_3$ formation. For these reasons, optimizing a direct thermal ALD process, without the above-mentioned seed-layers and pre-functionalization approaches, can be essential to achieve ultra-thin dielectric films on MoS$_2$, while preserving its structural integrity.

In the last years, different authors reported on the enhanced ALD nucleation of high-k dielectrics on monolayers (1L) of 2D materials, promoted by the interaction with the underlying substrate [17]. Such a behavior was first observed during thermal ALD of Al$_2$O$_3$ on CVD-grown graphene on copper [18], where the activation of the ALD process was attributed to peculiar polar sites produced by the graphene/metal interaction, acting as traps for the adsorption of the ALD-precursor [19,20]. An enhanced nucleation has been reported also in the early stages of thermal ALD of Al$_2$O$_3$ on top of 1L epitaxial graphene on 4H-SiC(0001), and it was ascribed to the peculiar interface structure of graphene with the SiC substrate [21,22]. This is characterized by the presence of an sp$^3$ carbon buffer layer, partially covalently bonded to the underlying Si face of SiC with a large density of positively charged dangling bonds, which are responsible of high n-type doping of epitaxial graphene. Ab-initio DFT calculations predicted an enhanced adsorption of the H$_2$O polar molecules, the co-reactants in the ALD process, on the highly doped graphene surface [22]. Furthermore, the compressive strain of epitaxial graphene on SiC was expected to significantly contribute to the increased reactivity with respect to an ideal free-standing graphene [22].

More recently, the direct growth of very uniform and ultra-thin (~4 nm) Al$_2$O$_3$ films with very good insulating properties has been achieved by thermal ALD at 250 °C upon 1L-MoS$_2$ membranes exfoliated on a gold substrate [17]. Interestingly, using identical ALD deposition conditions, an inhomogeneous Al$_2$O$_3$ coverage was observed for 1L-MoS$_2$ membranes produced by gold-assisted exfoliation and transferred onto an insulating Al$_2$O$_3$/Si substrate. The enhanced Al$_2$O$_3$ nucleation on the Au-supported 1L-MoS$_2$ was ascribed to an increased reactivity of MoS$_2$, originating from the strong S-Au interaction at MoS$_2$/gold interface [23,24]. However, the exact nanoscale mechanisms ruling the improved physisorption/chemisorption of the precursors and/or co-reactant molecules on the 1L-MoS$_2$/Au system during the early stages of the ALD process need to be further investigated, in view of future applications of this approach. Furthermore, it would be useful to understand whether the observed behavior is limited to the Al$_2$O$_3$ precursors or it can be extended to the ALD deposition of other high-k insulators, such as HfO$_2$, which are interesting for MoS$_2$ applications in electronics [25,26,27]. The commonly used Al precursor in the ALD growth of Al$_2$O$_3$, i.e. trimethylaluminum (TMA, Al(CH$_3$)$_3$) [28], is a relatively small molecule. On the other hand, the most frequently used Hf precursor for HfO$_2$ ALD growth, i.e. the tetrakis-dimethylaminohafnium (TDMAHf), exhibits a bulky structure, which can influence the nucleation degree due to steric hindrance [29].

In this paper, an insight on the possible mechanisms responsible of the enhanced reactivity of Au-supported 1L-MoS$_2$ towards the species involved in the thermal ALD process has been provided. The nanoscale structural/morphological properties and doping/strain for the experimental 1L-MoS$_2$/Au system has been initially assessed, and complemented by the predictions from ab-initio density functional theory (DFT) calculations for an ideal MoS$_2$/Au(111) heterojunction. Furthermore, the nucleation and growth of ultra-thin films of Al$_2$O$_3$ and HfO$_2$ on the surface of Au-supported 1L-MoS$_2$ under optimized ALD conditions was comparatively investigated by detailed atomic force microscopy investigations. By systematically following the surface coverage as a function of the number (N) of ALD cycles (N from 10 to 120), significant differences were observed between Al$_2$O$_3$ and HfO$_2$ (especially at lower N), and ascribed to the different nature of the used precursors. Finally, the insulating quality of uniform Al$_2$O$_3$ and HfO$_2$ films, obtained after 80 ALD cycles on 1L-MoS$_2$/Au, were assessed by nanoscale current mapping using conductive atomic force microscopy (CAFM), whereas their impact on the vibrational and light emission properties of MoS$_2$ has been investigated by Raman and Photoluminescence (PL) spectra.

## 2. Materials and Methods

The substrate material used in these experiments consisted of ~8 nm Ni wetting layer and ~10 nm Au film, sequentially deposited on a SiO$_2$/Si sample by DC magnetron sputtering. Therefore, a bulk molybdenite stamp was pressed on the freshly prepared Au surface, resulting in the effective exfoliation of ultra-thin and extended (cm$^2$) MoS$_2$ membranes, mostly formed by 1L-MoS$_2$ [23, 30,31,32,33,34,35,36].

Thermal ALD of Al$_2$O$_3$ and HfO$_2$ on the surface of the 1L-MoS$_2$/Au system was performed using in both cases H$_2$O as the co-reactant, while trimethylaluminum (TMA) and tetrakisdimethylaminohafnium (TDMAH) Hf [N(CH$_3$)$_2$]$_4$) were used as precursors for Al$_2$O$_3$ and HfO$_2$, respectively. During the Al$_2$O$_3$ process, the precursors pulse times of 20 ms were used for TMA and H$_2$O. Differently, for the HfO$_2$, the precursor pulse times were 40 ms and 20 ms for TDMAHf and H$_2$O, respectively. After each pulse, a purging N$_2$ was introduced in the chamber to remove unreacted precursors. The deposition temperatures, opportunely optimized for the two different precursors, were 250°C and 290°C for Al$_2$O$_3$ and HfO$_2$, respectively [37]. The pressure in the chamber during both processes was 10 Pa. The growth mechanism of the two high-κ dielectrics has been investigated on a set of consecutive ALD processes, increasing the number of deposition cycles, from 10 to 40, 80 and 120. The number of deposition cycles have been chosen identical for both dielectrics, considering the same growth rate of 0.88 Å/cycle calibrated on a silicon substrate.

The structural properties of the MoS$_2$ membranes exfoliated on the Au/Ni/SiO$_2$ substrate have been investigated with atomic resolution by electron microscopy on cross-sectioned lamellas, prepared by focused-ion-beam (FIB). First, high-resolution transmission electron microscopy (HRTEM), scanning transmission electron microscopy (STEM) in high-angle-annular-dark-field (HAADF) mode, and energy dispersive spectroscopy (EDS) analyses were acquired using an image corrected ThermoFisher THEMIS 200 microscope with 200 keV electron beam. Furthermore, atomic resolution analyses of the MoS$_2$/Au interface by STEM in the HAADF mode have been performed at primary electron beam energy of 200 keV using a probe corrected JEOL ARM 200F microscope.

The morphology of the as-exfoliated 1L-MoS$_2$ on gold, and the coverage degree of Al$_2$O$_3$ and HfO$_2$ on MoS$_2$ after different ALD cycles have been examined by tapping mode Atomic Force Microscopy (AFM), using a DI3100 equipment by Bruker with Nanoscope V electronics. Sharp silicon tips with a curvature radius of ~5 nm were used for these measurements. The insulating properties of the high-k films onto MoS$_2$/Au have been also evaluated by conductive-AFM (C-AFM) using the TUNA module by Bruker and Pt-coated Si tips.

Raman spectroscopy and PL measurements on the as-exfoliated MoS$_2$ and after the high-κ dielectric deposition have been executed by a Horiba HR-Evolution micro-Raman system with a confocal microscope (100× objective) and a laser excitation wavelength of 532 nm. A grating of 1800 lines/mm was employed for the Raman measurements in a range between 150-650 cm$^{-1}$, while a grating of 600 lines/mm was used for PL measurements in a range between 10-5500 cm$^{-1}$.

Finally, DFT calculations were performed in order to assess the structural and electronic properties of the MoS$_2$/Au(111) interface using the Quantum Espresso software package [38] with the van der Waals exchange-correlation functional of Hamada [39]. The modelled system comprised a single MoS$_2$ layer relaxing on an Au(111) bilayer. Due to the mismatch between the lattice parameters of the two parts of the heterostructure, a laterally strained Au(111) bilayer model was constructed in order to acquire the calculated equilibrium lattice parameters of the MoS$_2$ sheet (3.167 Å). Convergence was obtained with a plane-wave cut-off kinetic energy of 50 Ry and an augmented charge density cut-off of 400 Ry, using a (12×12×2) Monkhorst-Pack grid for the sampling of the Brillouin zone [40]. A vacuum space of 20 Å perpendicular to the MoS$_2$/Au(111) interface was inserted in order to avoid spurious interactions between the periodic replicas of the system. Calculations for the adsorption energy of ALD precursors (H$_2$O, TMA) on freestanding MoS$_2$ and the MoS$_2$/Au heterostructure were performed in (3×3) supercells with the same structural and convergence parameters as before, using a (4×4×2) Monkhorst-Pack grid for the sampling of the Brillouin zone.

## 3. Results and discussion
### 3.1 Nanoscale structure, strain and doping of 1L-MoS$_2$ exfoliated on gold

The morphological, structural and vibrational properties of the as-exfoliated MoS$_2$ membranes on the Au/Ni/SiO$_2$ stack were preliminary assessed by the combination of optical microscopy, AFM, micro-Raman and high resolution TEM/EDS analyses. Fig.1(a) shows an optical image of the sample surface, where 100 μm-wide regions covered by 1L-MoS$_2$ exhibit a slightly darker contrast as compared to bare Au regions. The presence of smaller size multilayer MoS$_2$ areas, with a dark violet contrast can be also observed in the image. Fig.1(b) shows a high-magnification AFM morphology collected at the boundary region between the 1L-MoS$_2$/Au and a bare Au area. The roughness of the Au surface is due to the grain structure of the metal film deposited by sputtering, and the MoS$_2$ membrane seems to be very conformal to the morphology of underlying Au. To obtain quantitative information on the thickness and root mean square roughness of 1L-MoS$_2$ with respect to the Au surface, the histogram of the height values was extracted from this morphological analysis and reported in Fig.1(c). This distribution exhibits two well separated peaks, associated to the bare Au and 1L MoS$_2$/Au areas, which could be fitted by Gaussian contributions with very similar standard deviations ($\sigma_{Au}$=0.25 nm and $\sigma_{MoS2/Au}$=0.28 nm), indicating a conformal coverage of MoS$_2$ on Au. Furthermore the separation between these peaks (t=0.64 nm) is very close to the typically reported thickness of 1L-MoS$_2$ evaluated by AFM analyses [17].

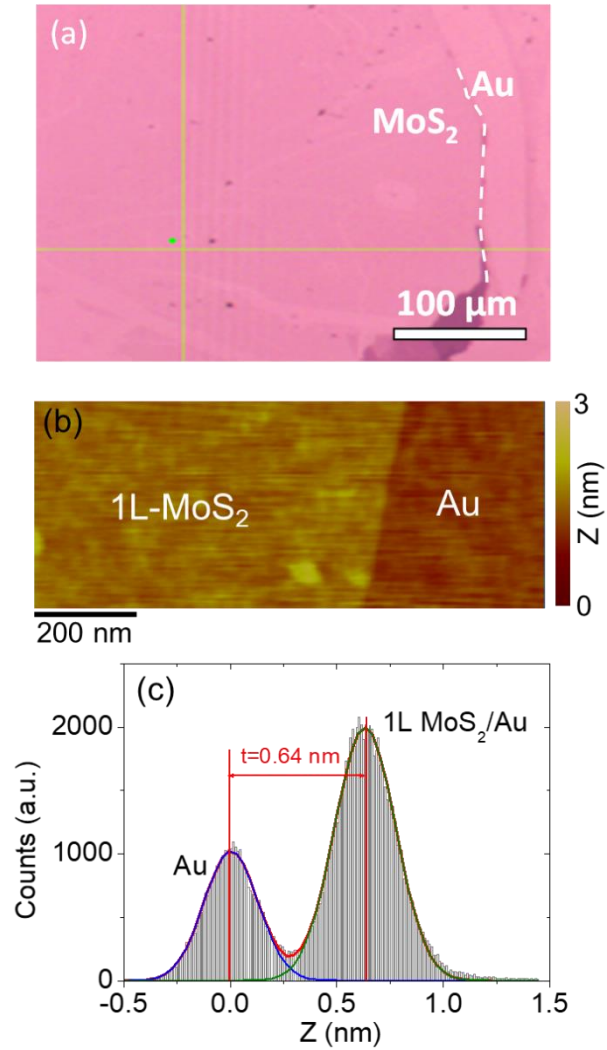

**Figure 1** (a) Optical image of the MoS$_2$ membrane exfoliated on Au, where the brighter contrast correspond to bare Au regions, a slightly darker contrast to extended areas covered by 1L-MoS$_2$, and the dark violet contrast to areas with multilayer MoS$_2$. (b) AFM topography in a region at the boundary between 1L-MoS$_2$/Au and bare Au. (c) Histogram of the height values extracted from the AFM image, and fit with two Gaussian contributions. The root mean square roughness of the two regions and the thickness of 1L-MoS$_2$ (t=0.64 nm) with respect to the Au substrate were evaluated from this analysis.

Fig.2(a) shows a typical Raman spectrum collected on the 1L-MoS$_2$ membrane on Au. Noteworthy, the measured separation $\Delta\omega \approx 21.2$ cm$^{-1}$ between the out-of-plane (A$_{1g}$) and in-plane (E$_{2g}$) vibrational modes is larger than the typically reported values (from 18 to 19 cm$^{-1}$) for 1L-MoS$_2$ on commonly used insulating substrates (such as SiO$_2$) [41]. This discrepancy for gold-supported 1L-MoS$_2$ has been attributed to the combined effect of a tensile strain (causing a red-shift of the E$_{2g}$ peak) and of a p-type doping (causing a blue-shift of the A$_{1g}$ peak) [23,42].

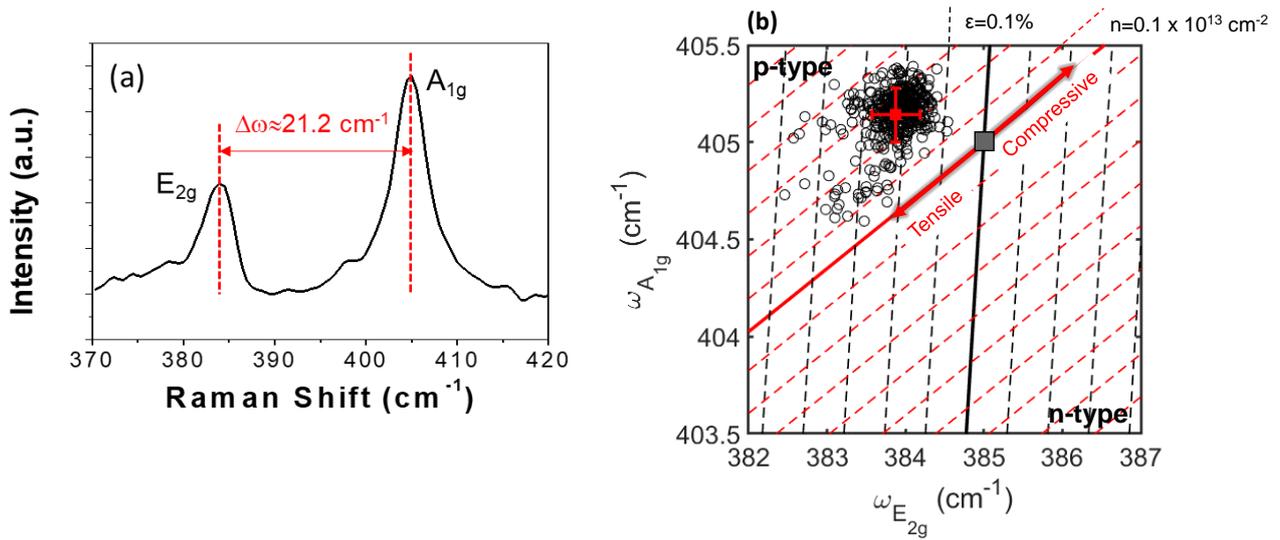

**Figure 2** (a) Typical Raman spectrum of 1L-MoS$_2$ exfoliated on Au, with the indication of the characteristic in-plane (E$_{2g}$) and out-of-plane (A$_{1g}$) vibrational modes and their separation $\Delta\omega \approx 21.2$ cm$^{-1}$. (b) Correlative plot of the $\omega_{A1g}$ and $\omega_{E2g}$ peak frequencies extracted from an array of Raman spectra measured at different positions on the 1L-MoS$_2$/Au sample.

The Raman spectra reported in Fig.2(a) are similar to the ones reported in some recent papers [30,32,33,36]. On the other hand, they differ from the spectra reported in other works [34,35], which showed an anomalous splitting in two components of the A$_{1g}$ peak (mainly related to MoS$_2$ doping), and a large redshift of the E$_{2g}$ peak (mainly related to MoS$_2$ strain). The A$_{1g}$ peak splitting was ascribed in Ref. [34] to a non-conformal morphology of the exfoliated MoS$_2$ on the Au surface, with a large density of nanoscale regions where the membrane is suspended (i.e. not directly in contact with the Au surface). This resulted in the coexistence of doped and undoped MoS$_2$ patches, which are reflected in the two components of the A$_{1g}$ peak. The large redshift of the E$_{2g}$ peak is due to the large strain of MoS$_2$, probably related to this peculiar morphology. Hence, the main differences between our MoS$_2$/Au samples and those investigated in these recent literature works arise from a more conformal morphology of MoS$_2$ on the Au surface.

The MoS$_2$ strain and doping can be quantitatively evaluated by a correlative plot of the A$_{1g}$ vs E$_{2g}$ wave-numbers, as illustrated in Fig.2(b). Here, the open points represent the experimental ($\omega_{E2g}$, $\omega_{A1g}$) values measures on an array of 21×21 laser positions in a 5μm×5μm area of 1L-MoS$_2$ surface, whereas the red (black) continuous lines are strain (doping) lines, illustrating the theoretical relations in the case of an ideally undoped (unstrained) monolayer of MoS$_2$. The literature values of the E$_{2g}$ and A$_{1g}$ peaks' frequencies measured on a suspended monolayer MoS$_2$ ($\omega^0_{E2g}$=385 cm$^{-1}$, $\omega^0_{A1g}$=405 cm$^{-1}$) [43] were taken as the origin in the plot. In fact, freestanding MoS$_2$ can be considered as a good approximation of the ideal unstrained and undoped monolayer, since the effects of the interaction with the substrate are removed. From this plot, a tensile strain ranging from ε≈-0.1% to ε≈-0.3%, and a p-type doping ranging from $0.1\times10^{13}$ to $0.4\times10^{13}$ cm$^{-2}$ can be observed. Obviously, these are averaged values on the micrometer area probed by the laser spot. It is expected that local values of strain exhibit larger variations at the nanoscale, depending on the local morphological/structural properties of MoS$_2$/Au interface.

To get a deeper insight on these aspects, the structural properties of the MoS$_2$/Au heterointerface have been investigated by cross-sectional TEM/STEM analyses. Fig.3(a) shows a low-magnification TEM analysis of the whole MoS$_2$/Au/Ni/SiO$_2$ stack, from which the thickness and the grain size of the Ni wetting layer and of the Au films can be directly deduced. The presence of the 1L-MoS$_2$ on the Au surface can be also observed in this image, and it is further confirmed by the high magnification EDS chemical maps of Mo, S and Au distributions, reported in Fig.3(b), (c) and (d), respectively. Interestingly, such nanoscale analysis confirms that the ultra-thin MoS$_2$ membrane generally follows the morphology of Au grains, appearing to be suspended in some regions, typically corresponding to Au grain boundaries (see white arrow in Fig.3(a)).

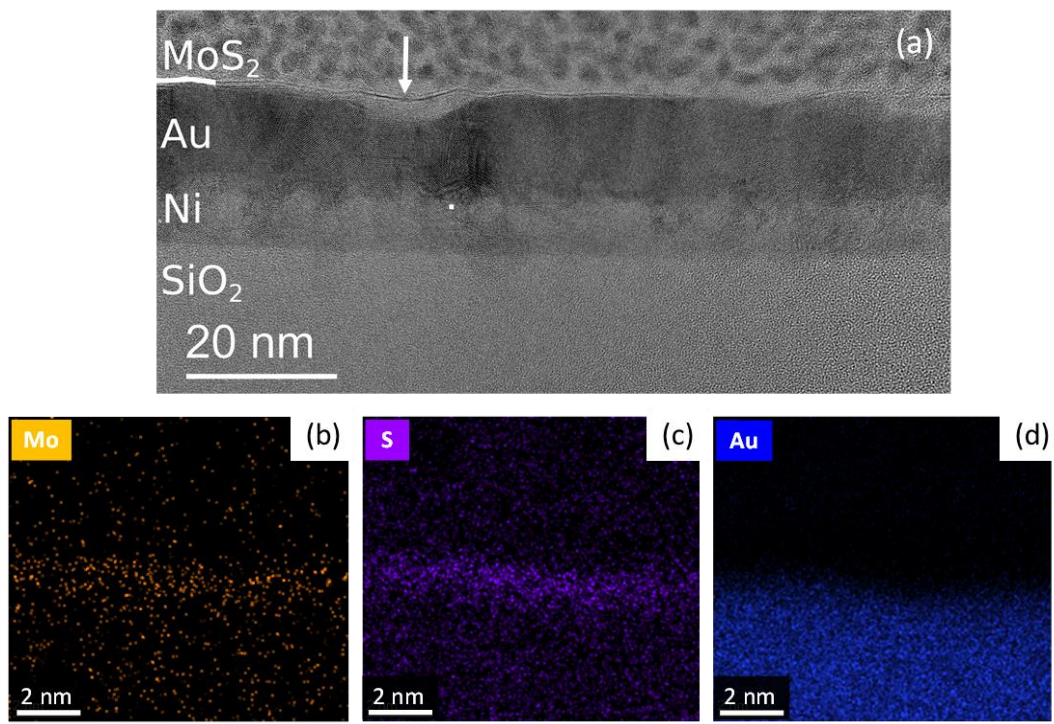

**Figure 3**: (a) Low magnification TEM analysis of the monolayer MoS$_2$ membrane on the Au/Ni/SiO$_2$ stack. EDS maps of Mo (b), S (c), Au (d) acquired on the magnified MoS$_2$/Au region.

Atomic resolution information on the 1L-MoS$_2$/Au interface structure were obtained by aberration corrected STEM analyses in the HAADF mode. In this imaging mode, based on the collection of electrons scattered at high angle, atomic columns of elements with high atomic number Z, such as Au and Mo exhibit a brighter contrast. Fig.4(a) and (c) show two representative analyses showing the MoS$_2$ residing on the atomic terraces and steps exposed by the gold grains.

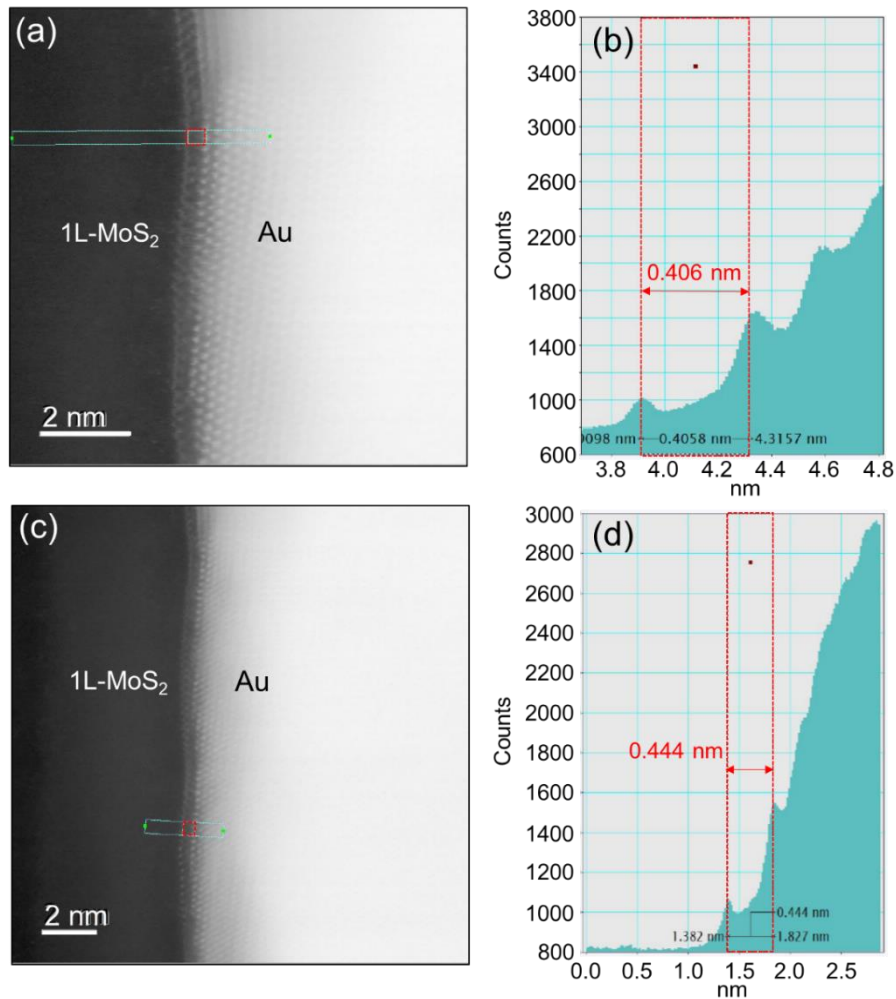

**Figure 4**: (a,c) Aberration corrected STEEM/HAADF analyses of 1L-MoS$_2$/Au interface. (b,d) Intensity line profiles, from which the atomic distance between the Mo atoms of the MoS$_2$ membrane and the topmost Au atoms was evaluated.

The atomic distances ($d_{Mo-Au}$) between the Mo atoms of the MoS$_2$ layer and the topmost Au atoms was evaluated from the intensity linescans, as illustrated in Fig.4(b) and (d). A variability of $d_{Mo-Au}$ from ~4.0±0.1 Å to ~4.5±0.1 Å, depending on the local configuration of Au planes, is observed, which correspond to S-Au interatomic distances ($d_{S-Au}$) of 2.4-2.9 Å. These values are indeed lower than the S-Au distance of 3.5 Å obtained by Velický et al. [32], starting from a $d_{Mo-Au}$ value of 5.1±0.3 Å, measured by STEM. The origin of the smaller $d_{Mo-Au}$ distances obtained in our work may be related to a more conformal adhesion of MoS$_2$ to the Au surface, as confirmed by AFM morphologies in Fig.1(b) and (c), showing similar roughness values on $MoS_2$/Au and bare Au surfaces, and by Raman spectra in Fig.2(a), without splitting of the $A_{1g}$ peak. We want also to point out that, although the $d_{S-Au}$ distances obtained in our work indicate a stronger S-Au interaction than a common vdW bond, they are still higher than the expected $d_{S-Au}$ distance (2.2 Å) for a S-Au covalent bond [32]. Furthermore, these $d_{S-Au}$ distances are lower than the interlayer vdW spacing $d_{S-S}$ of 3.0 Å between S atoms of $MoS_2$ layers in bulk $MoS_2$, which is reasonable to explain monolayer $MoS_2$ exfoliation of Au.

### 3.2 DFT calculations of an ideal $MoS_2$/Au system and its reactivity to ALD precursors

In order to obtain a better insight on the fundamental structural/electronic properties of the $MoS_2$/Au(111) system, we have computed the equilibrium distance and binding energy of this interface using DFT calculations. Upon structural relaxation, the coupled system largely preserved the structural characteristics of the respective pristine structures, indicating a van der Waals heterostructure (see Fig. 5). The obtained Mo-Au equilibrium distance of 4.28 Å is in line with the experimental measurements, whereas a strong binding energy per interface atom was calculated (~0.26 eV), implying a rather strong attachment for a van der Waals system. This aspect explains the successful exfoliation of ultra-thin $MoS_2$ membranes through pressing of molybdenite stamps on the Au surface and the resulting acquisition of the Au surface morphology from the $MoS_2$ layers. Electronically (Fig. 5a), the interface is characterized by a finite density of states around the Fermi level due to the presence of states deriving from the Au substrate. On the other hand, the $MoS_2$ layer preserves the direct nature of its bandgap for monolayer structures at the *K* Brillouin zone point (Fig. 5b), showing however a partial hybridization of its Mo 4d orbitals with Au(111) surface states.

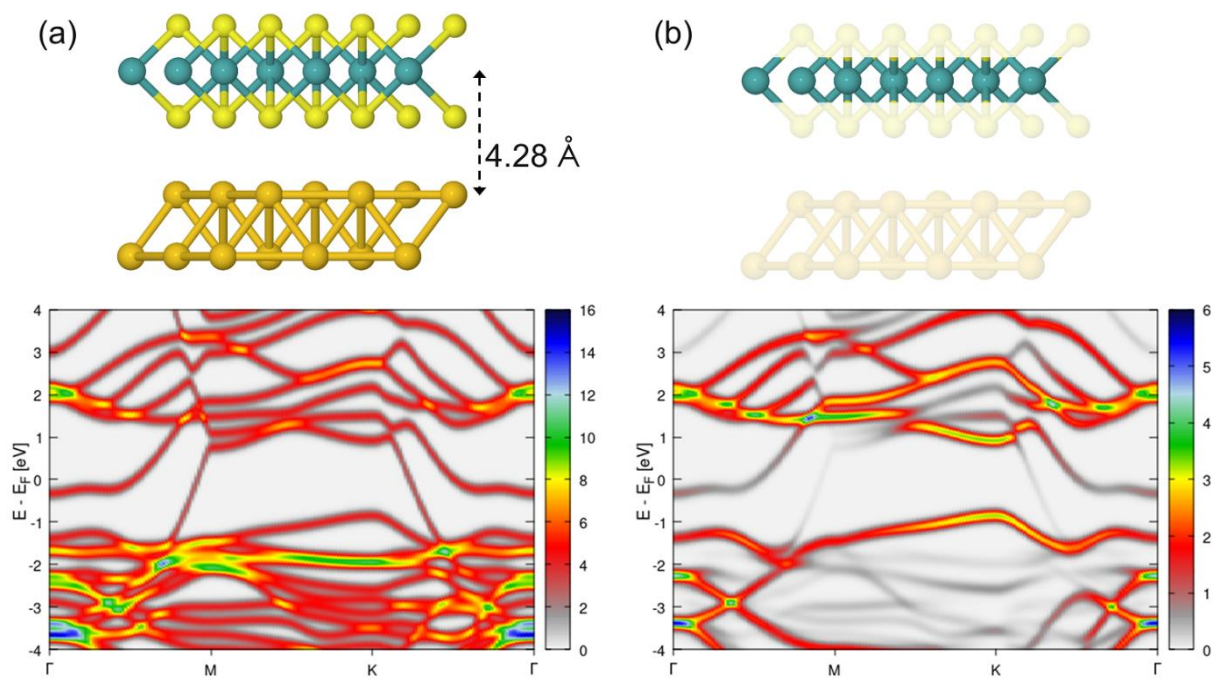

**Figure 5:** Structure and electronic properties of the MoS$_2$/Au(111) interface showing the (a) the total k-resolved projected density of states, and (b) the partial k-resolved projected density of states only for Mo 4d orbitals.

We have thereon focused on the ALD deposition process and calculated the adsorption energy of all ALD precursors (H$_2$O, TMA, TDMAHf) on the MoS$_2$/Au system, comparing these values with respective ones obtained for the interaction between the same molecules and freestanding MoS$_2$. Fig.S1 of the Supporting Information shows the optimized configurations for the adsorption of (a) H$_2$O, (b) TMA (monomer), and (c) TDMAH molecules on the MoS2/Au(111) surface, based on DFT calculations. The results of this simplified model for the 1L-MoS$_2$/Au system showed that the Au substrate slightly improves the adsorption of the TDMAHf precursor on MoS$_2$ but only marginally affects the adsorption of the H$_2$O and TMA molecular species (see Tab.1). This indicates that the enhanced ALD nucleation on Au-supported MoS$_2$ should be mainly related to morphological (e.g., surface curvature, bound and unbound MoS$_2$ regions) and associated structural modifications (e.g., strain or deformation of the MoS$_2$ layer), as shown by the previous characterizations of the MoS$_2$/Au system. To this end, a detailed experimental analysis on the nucleation process should better clarify the studied ALD growth mechanism.

**Table 1:** Adsorption energy (in eV) calculated for an $H_2O$, TMA (monomer) and TDMAHf molecule on free-standing $MoS_2$ and $MoS_2$ supported on an Au(111) substrate.

|  | $H_2O$ | TMA (monomer) | TDMAHf |
|---|---|---|---|
| $MoS_2$ | -0.11 | -0.47 | -0.46 |
| Au(111)/$MoS_2$ | -0.12 | -0.48 | -0.52 |

### 3.3 Nucleation and growth of $Al_2O_3$ and $HfO_2$ on $MoS_2$/Au

In this section, the nucleation and growth of $HfO_2$ and $Al_2O_3$ on the $MoS_2$/Au system has been investigated for different number of ALD cycles. The coverage degree of the $MoS_2$ surface by both high-κ dielectrics has been evaluated by tapping mode AFM. Fig.6 displays the comparison of AFM-morphology (a) and phase (b) of 40 ALD cycles $HfO_2$ on $MoS_2$/Au, chosen as representative sample. The morphological map indicates an incomplete coverage of $MoS_2$ surface by the growing $HfO_2$ grains. This is further confirmed by the phase map, which is known to be sensitive to the surface composition. In particular, the violet color in the phase image coincides with $HfO_2$ covered regions, as shown in the morphological image, whereas the green phase contrast is associated to the uncovered areas. In the following, the AFM phase maps have been used to evaluate the ALD coverage degree as a function of the number of cycles for both $Al_2O_3$ and $HfO_2$.

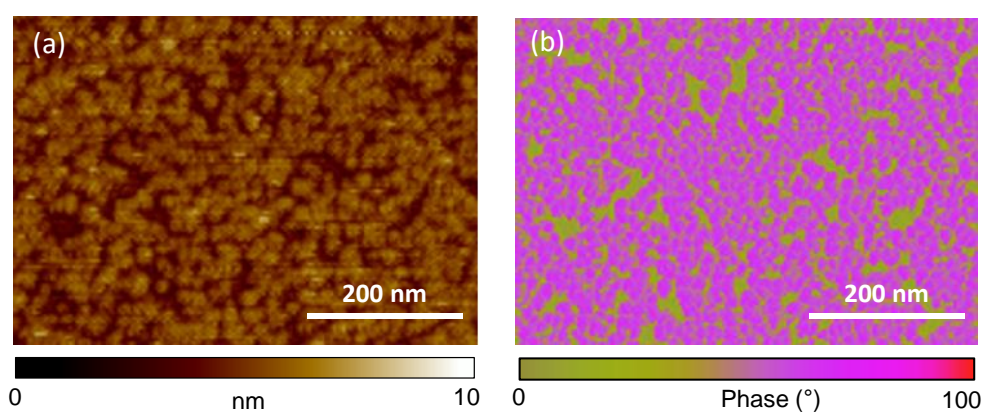

**Figure 6**: Comparison of AFM morphology (a) and phase (b) of 40 ALD cycles $HfO_2$ on $MoS_2$/Au

Fig. 7 compares the phase maps of $Al_2O_3$ (a) and of $HfO_2$ (b) grown on 1L- $MoS_2$ after 10, 40, 80 and 120 ALD cycles. As previously noted, the violet and green regions indicate the covered and uncovered MoS$_2$ surface by the high-κ dielectrics, respectively. The evolution of surface percentage coverage, calculated from the phase maps, is reported in Fig. 7(c). It can be observed how, after only 10 cycles, Al$_2$O$_3$ covers 50 ± 3 % of the MoS$_2$ surface, whereas it reaches 93 ± 2 % coverage after 40 cycles, and 100% after 80 and 120 cycles, as reported in our previous work [17]. Differently, the growth evolution of HfO$_2$ appears to be slower during the initial deposition cycles. In particular, it exhibits a coverage degree of about 15 ± 3 % after 10 cycles and 70 ± 3 % after 40 cycles. An almost complete layer (97 ± 3 %) is obtained after 80 cycles and a 100 % coverage after 120 cycles.

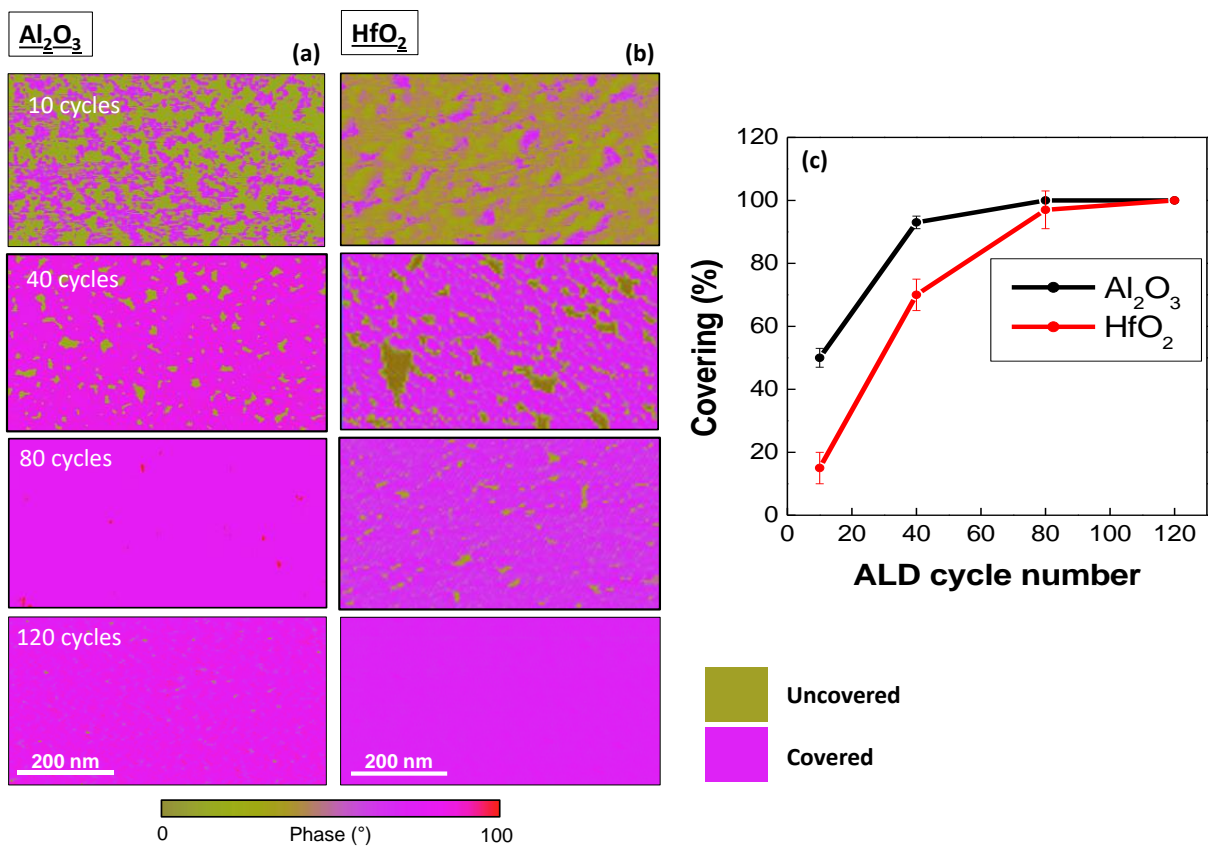

**Figure 7**: AFM phase maps afer 10, 40, 80 and 120 ALD cycles of deposition Al$_2$O$_3$ (a) and HfO$_2$ (b) on monolayer-MoS$_2$/Gold. The green contrast represents the uncovered MoS$_2$/Gold by high-κ dielectrics. Conversely, the magenta contrast represents the covered MoS$_2$/Gold by high-κ dielectrics. The coverage degree between the Al$_2$O$_3$ and HfO$_2$ ALD deposition is graphically compared in (c).

To get further insight in the nucleation mechanism of the two high-κ insulators on the Au-supported 1L-MoS$_2$, the grain size distributions in the early stages of ALD deposition have been investigated in detail. Fig.8(a) and (d) show the AFM morphologies of Al$_2$O$_3$ and HfO$_2$ after the 40-cycles ALD processes. To precisely evaluate the island sizes, the grain boundaries have been highlighted by image processing, as displayed in Fig.8(b) and (e). Therefore, the histograms of grain diameters have been evaluated, as shown in Fig.8(c) and (f).

Differently than the layer-by-layer ALD growth, typically reported on common semiconductor substrates and resulting in the formation of conformal and flat dielectric films, a 3D-growth mode is observed on the surface of van der Waals layered materials, due to the weak interaction between precursors and substrate. In this case the growth process evolves by the initial islands formation and their subsequent coalescence toward a full layer. In this regard, recent works have proposed some models to describe 3D-growth mechanisms depending on the main deposition parameters, the nature of deposited material and substrate, and their interactions [44,45]. In general, these models take into account: (i) the cyclic deposition of atoms on the surface of the substrate and of the islands formed during previous ALD cycles; (ii) the diffusion/aggregation of the deposited atoms on the substrate to form new islands; (iii) the diffusion and coalescence of islands (dynamic coalescence). In this context, Grillo et al. [44,45] showed how the peculiar shape of the island-size distribution measured at a certain stage of the ALD process provides information on the specific growth mechanism.

Hence, we analyzed the island maps and the size distributions, reported in Fig.8, to draw some conclusion about the growth mechanism of $Al_2O_3$ and $HfO_2$ on the 1L-$MoS_2$/Au system. We found that the $Al_2O_3$ morphology (Fig.8 (a)) is formed by nanometric islands, densely arranged, and with a modest uncovered portion of $MoS_2$ (~ 7 %) as shown in Fig.8 (b). Moreover, the distribution of $Al_2O_3$ island's diameters, reported in Fig.8(c), exhibits a bimodal shape fitted with two Gaussian contributions, i.e. a narrow peak at ~7 nm and a broader peak at ~12 nm. Differently, the $HfO_2$ morphology is formed by larger islands and a larger uncovered fraction of $MoS_2$ (30 %), as shown in Fig. 8 (d), (e). The distribution of $HfO_2$ island's diameters (Fig. 8 (f)) also exhibits a bimodal shape, with a narrow peak at ~7 nm and a broader peak at ~20 nm. The small-size peak is attributable to the new atoms arriving on the surface of the substrate. The large-size peak is correlated to the growing islands as a result of new deposition, diffusion, and coalescence. Hence, the bimodal shape, obtained for both high-κ, is an indication of a nucleation mechanism involving a cyclic deposition of materials both on bare substrate and on preexisting islands, as well as a dynamic coalescence. However, $HfO_2$ exhibits a broader large-size peak, shifted toward 20 nm, than $Al_2O_3$. The formation of larger grains can be correlated to a lower diffusivity in the case of $HfO_2$ due to its bulky metalorganic precursor. Since diffusion processes help the 2D-growth, when they are disadvantaged, a high number of cycles are required to reach the complete covering. This mechanism can be the explanation of the delay of $HfO_2$ process than $Al_2O_3$. Hence, although DFT calculations predicted a slightly better adsorption of single TDMAHf molecules on MoS2/Au (Table 1), the steric hindrance effect due to bulky structure of this precursor represents the limiting factor for HfO$_2$ nucleation.

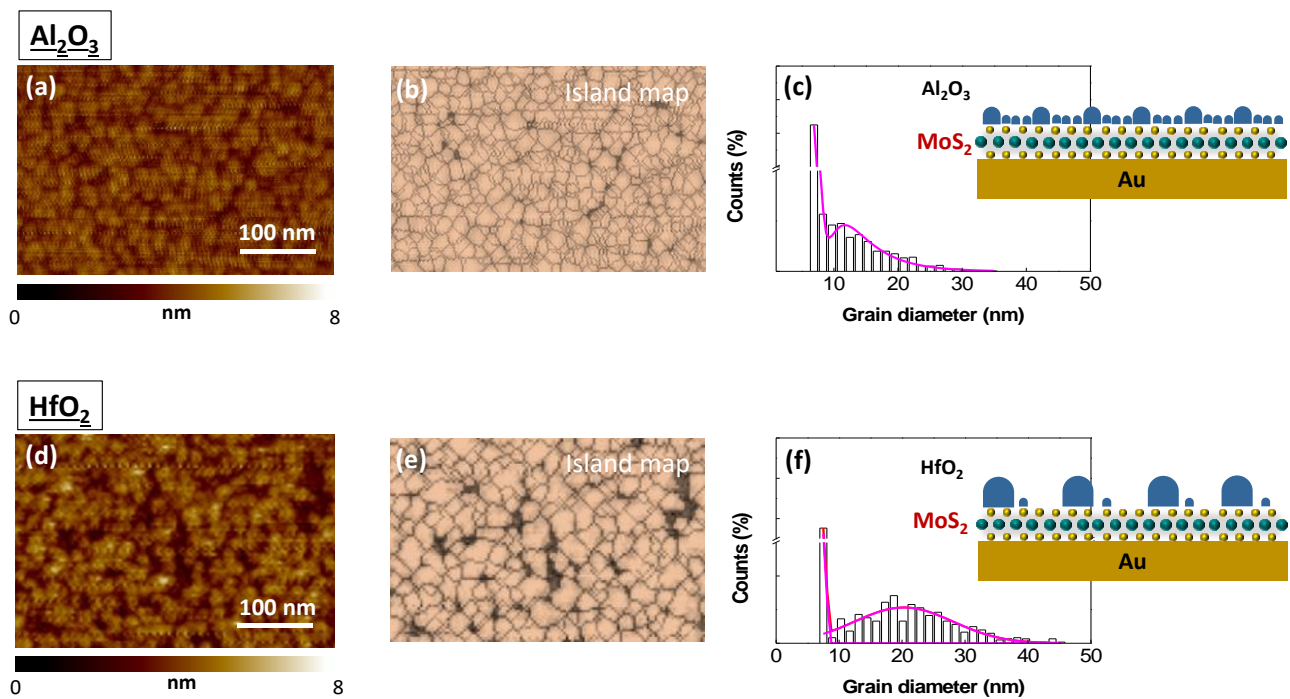

**Figure 8**: High-magnification AFM morphology of Al$_2$O$_3$/MoS$_2$ (a) and HfO$_2$/MoS$_2$ (d) after 40 ALD cycles. Island maps of Al$_2$O$_3$/MoS$_2$ (b) and HfO$_2$/MoS$_2$ (d) after 40 ALD cycles reproduced by the AFM morphologies (a) and (b). Distributions of the islands diameters for Al$_2$O$_3$/MoS$_2$ (c) and HfO$_2$/MoS$_2$ (f) referred to the islands map of (b) and (e). The schematics superimposed on (c) and (f) are illustrates the nucleation mechanism of Al$_2$O$_3$ and HfO$_2$ on MoS$_2$, respectively.

### 3.4 Insulating properties of the high-k layers

The electrical quality of the high-κ layers has been further evaluated by conductive-AFM analyses. In particular, by scanning the metal-coated tip across the step between the high-κ grown on MoS$_2$ and the underlying Au substrate (as illustrated in Fig.9 (a)), we acquired the topographies (Fig.9 (b), (d)) as well as the current maps of the same region (Fig.9 (c), (e)). From the measured steps in the height line-scans of the Al$_2$O$_3$/MoS$_2$/Au (insert of Fig. 9 (b)) and HfO$_2$/MoS$_2$/Au (insert of Fig. 9 (d)) and after subtracting the monolayer MoS$_2$ thickness (~0.64 nm), the thicknesses of the two dielectrics after 80 ALD cycles were estimated as ~3.6 nm and ~3.1 nm, respectively. These measured thickness were about half those expected based on the growth rate previously evaluated on Si (0.88 Å/cycle). This clearly indicates that the ALD growth on the MoS$_2$/Au surface is much slower than on Si, especially in the early stages of the deposition. Of course, this is a consequence of the lack of out-of-plane bonds (which act as active sites of nucleation) on the MoS$_2$ surface, differently than on the Si surface. Despite this delayed growth, it is important noting that a uniform nucleation is obtained on

Au-supported MoS$_2$, resulting in the formation of homogeneous dielectric films after 80 cycles. A completely different scenario, i.e. a very inhomogeneous dielectric film, is typically observed for similar deposition conditions on MoS$_2$ supported by an insulating substrate (such as SiO$_2$ or Al$_2$O$_3$ on silicon) [17].

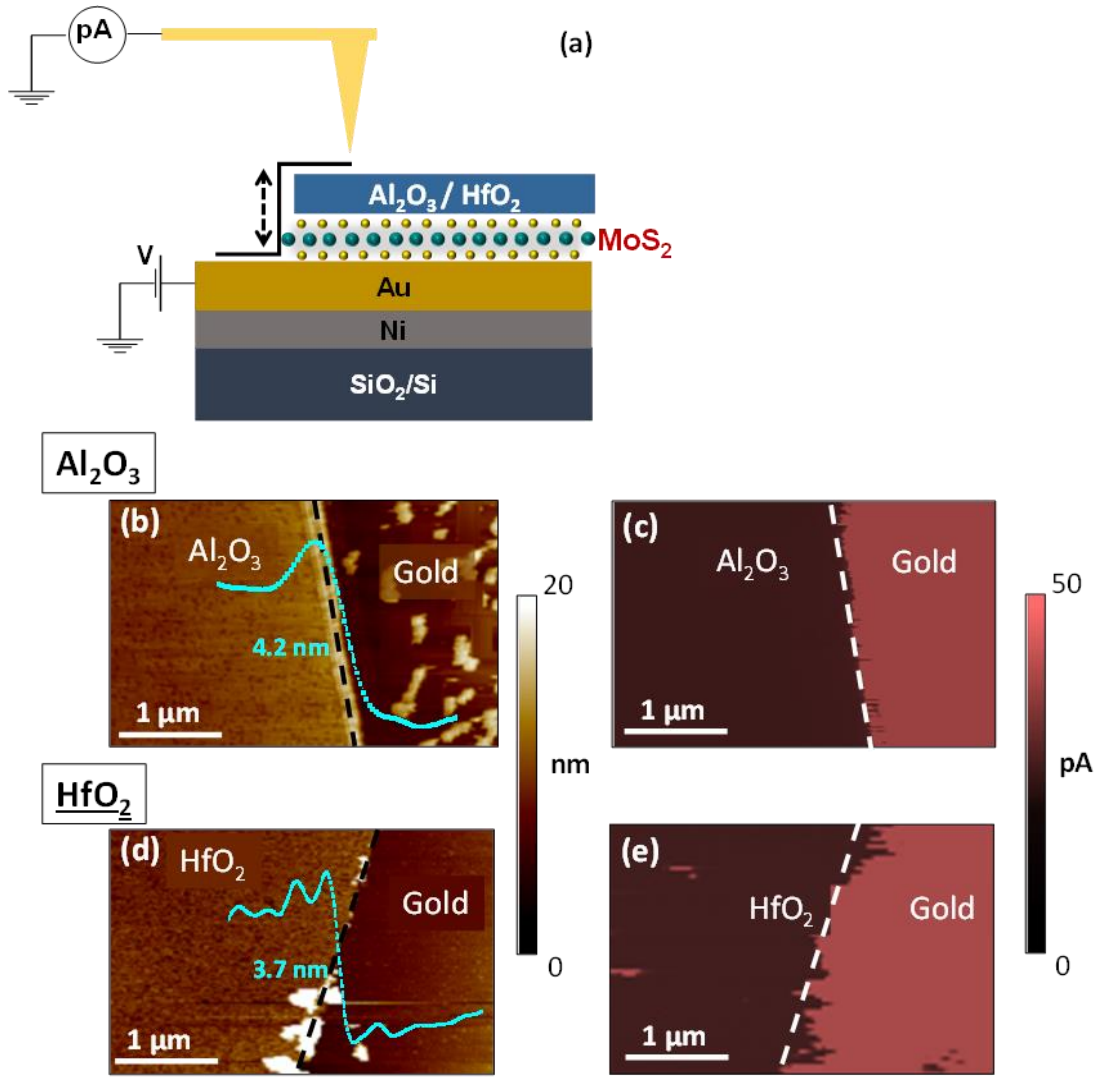

**Figure 9**: Schematic of the step between the high-κ (Al$_2$O$_3$ or HfO$_2$)/ MoS$_2$ and the underlying gold and with the circuit used for conductive-AFM measurements (a). AFM image of the step between Al$_2$O$_3$/MoS$_2$ (b) and HfO$_2$/MoS$_2$ (e) on Au after 80 ALD cycles. Height line-scans of the Al$_2$O$_3$/MoS$_2$/Au and HfO$_2$/MoS$_2$/Au from which the thichness of the dielectrics have been determinated are reported as insert in (b) and (c), respectively. Current maps of the step between Al$_2$O$_3$/MoS$_2$ (c) and HfO$_2$/MoS$_2$ (e) on Au after 80 ALD cycles, acquired on the same regions of (b) and (d), with a bias V= 3 V.

By applying a bias of 3 V between the tip and the gold substrate, it is possible to observe in Fig.9 (c) a very homogeneous insulating behavior for the Al$_2$O$_3$ on MoS$_2$, and the expected conductive behavior for gold. Under the same applied bias conditions, the HfO$_2$ layer grown on MoS$_2$ (Fig.9 (f)) exhibits good insulating performances, with the presence of some isolated current spots, which can be associated with less dense regions of the HfO$_2$ layer. This phenomenon is observed mainly on the HfO$_2$ layer probably due to a more inhomogeneous nucleation than the Al$_2$O$_3$ during the early stages. A current map has been also acquired at the same bias of 3V on the HfO$_2$ film deposited on MoS$_2$/Au by 120 ALD cycles. As shown in Fig.S2(a) of the Supporting Information, uniform insulating properties (without current spots) and a lower current level is observed, due to the higher thickness (~ 4.7 nm) of the HfO$_2$ film.

### 3.5 Impact of Al$_2$O$_3$ and HfO$_2$ deposition on the strain, doping and optical emission of 1L-MoS$_2$

Micro-Raman spectroscopy analyses have been performed to investigate the effect, in terms of strain and doping, of the high-k deposition on 1L-MoS$_2$. In order to correctly estimate these effects, Raman characterization has been performed on the samples that present a complete dielectric layer. Thus, Al$_2$O$_3$ and HfO$_2$ layers grown using 80 cycles have been investigated.

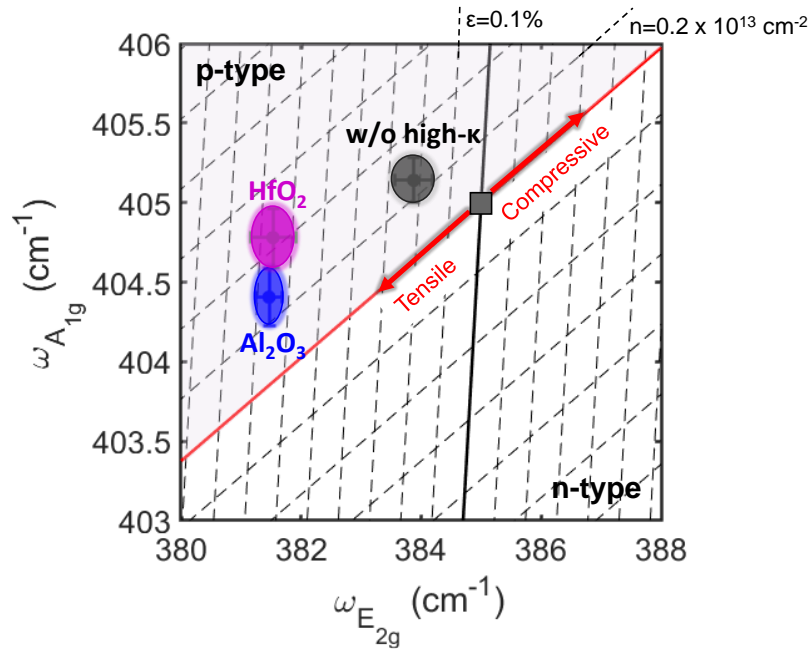

**Figure 10**: Correlative plot of the A$_1$' versus E' of the Raman spectra acquired on monolayer MoS$_2$ on Gold without high-κ (black circle) and of the same covered with 80 ALD cycles of Al$_2$O$_3$ (blue circle) and HfO$_2$ (magenta circle), allowing to estimate the type and the average values of strain and doping of monolayer MoS$_2$. The spacing of the dashed lines, parallel to the ideal strain and doping lines, is associated with carrier density changes of $0.2 \times 10^{13}$ cm$^{-2}$ and strain changes of 0.1%, respectively.

In Fig.10, the strain and doping of MoS$_2$ before and after the ALD processes have been estimated by the correlative analysis of A$_{1g}$ versus E$_{2g}$ peak frequencies, similarly to Fig.2(b). In Fig.10, the black circle indicates the range of experimental data points measured on 1L-MoS$_2$/Au before the ALD processes. It is characterized by an average tensile strain of ε≈0.21% and p-type doping of $0.25 \times 10^{13}$ $cm^{-2}$. Differently, the blue ellipse, referred to $Al_2O_3/MoS_2$ system, indicates an average tensile strain of ε≈0.65% and p-type doping of $0.25 \times 10^{13}$ $cm^{-2}$. While, the magenta ellipse, related to $HfO_2/MoS_2$ sample, exhibits an average tensile strain of ε≈ 0.65% and p-type doping of $0.45 \times 10^{13}$ $cm^{-2}$. It is obvious the strong impact of both high-k on strain properties of $MoS_2$, which changes from ε≈ 0.21% to ε≈ 0.65%, before and after the ALD process, respectively. Moreover, $Al_2O_3$ layer causes a negligible effect on the $MoS_2$ doping, which remains unchanged ($0.25 \times 10^{13}$ $cm^{-2}$). Conversely, $HfO_2$ is responsible of a higher p-type doping of $0.45 \times 10^{13}$ $cm^{-2}$. This higher doping can be due to a higher density of charged defects at the $HfO_2/MoS_2$ interface, due to the less efficient nucleation during the early growth stage of $HfO_2$ than $Al_2O_3$.

In the final part of this paper, the effects of $Al_2O_3$ and $HfO_2$ deposition on the photoluminescence (PL) spectra of 1L-$MoS_2$ is investigated. The interaction between $MoS_2$ and the Au substrate is well known to cause a total [34] or partial quenching [33,36] of the PL of monolayer $MoS_2$, as compared to $MoS_2$ residing on insulating substrates. In our previous papers [23,42] on 1L-$MoS_2$/Au samples, we have observed that the PL intensity is significantly quenched (but not totally suppressed), and the main PL peak is red-shifted compared to 1L-$MoS_2$ on an $Al_2O_3$ substrate. The origin of PL quenching in the 1L $MoS_2$/Au system can be ascribed to different factors, such as the preferential transfer of photoexcited charges to the Au substrate, and a large strain of the $MoS_2$ membrane causing a modification of the energy bandstructure [46]. In our case, the 1L-$MoS_2$ is nearly conformal to the underlying Au, but its strain is ε≈-0.3%, as shown in Fig.2(b) Probably this allows us to measure a PL spectrum, although partially quenched.

In Fig. 11, the black dots represent the PL spectra of bare monolayer $MoS_2$ on Au (a) and after 80 ALD cycles, resulting in the formation of continuous films of $Al_2O_3$ (b) and $HfO_2$ (c). A significant reduction (from ~2-4 times) of the main PL peak intensity and a modification of the shape of the spectra can be observed after $MoS_2$ encapsulation with $Al_2O_3$ and $HfO_2$. Such lowering of the PL intensity can be correlated, in the first approximation, to the increase of tensile strain in the $MoS_2$ membrane observed by Raman data in Fig.10, coherently with previous reports [17,43]. To get further insights on the differences between the $Al_2O_3$ and $HfO_2$ coated samples with respect to the bare 1L-$MoS_2$/Au, a deconvolution analysis of the PL spectra was carried out in Fig.11(a)-(c). All the experimental data can be well fitted by three different peaks, associated to the excitonic contributions A (blue line) and B (magenta line), and to the trionic contribution $A^-$ (green line), respectively [23]. This analysis revealed a significant reduction of the intensity ratio between the trionic ($A^-$) and the excitonic (A) peaks, from ~1.8 on the bare 1L-$MoS_2$ to ~0.8 and ~0.9, after the deposition of $Al_2O_3$ and $HfO_2$, res pectively. Recently, Lin et al. [47] demonstrated a decreasing trend of the A$^-$/A intensity ratio in the PL spectra of 1L-MoS$_2$ with the effective dielectric constant of the surrounding environment. Hence, in addition to overall quenching of the PL signal due to the increased compressive strain, the encapsulation of 1L-MoS$_2$ with the high-k dielectric films modifies the relative intensities of A$^-$ and A contributions, resulting in a change in the shape of the PL spectra.

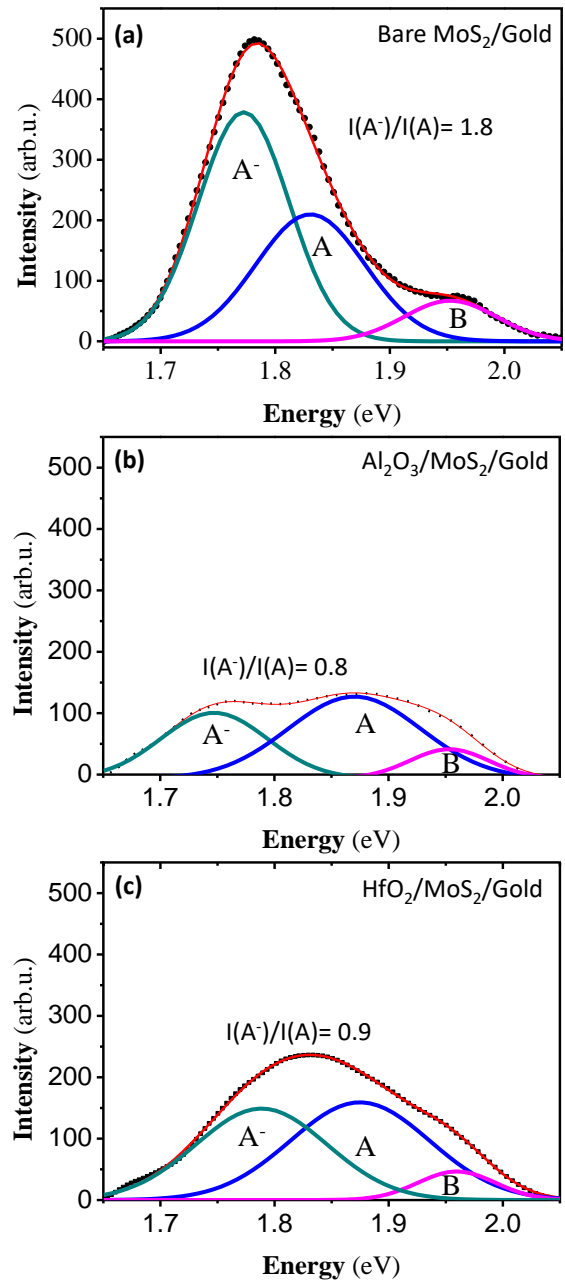

**Figure 11**: Micro-PL spectra on the bare 1L-MoS$_2$/Au (a) and after 80 ALD cycles of Al$_2$O$_3$ (b) and HfO$_2$ (c). The deconvolution of the PL peak results in three different contributions: the trion peak A$^-$, the exciton A, the exciton B.

## 4. Conclusions

In conclusion, we have demonstrated the direct thermal ALD of two technologically relevant high-k insulators ($Al_2O_3$ and $HfO_2$) on gold-supported monolayer (1L) $MoS_2$, providing an insight on the mechanisms ruling the nucleation in the early stages of the ALD process. The strain, doping, and morphology of the large area 1L-$MoS_2$ exfoliated on sputter-grown Au/Ni films have been preliminarily assessed by a complete multiscale characterization. Micro-Raman mapping demonstrated a tensile strain and p-type doping distribution on $MoS_2$, induced by the interaction with the substrate. Nanoscale resolution morphological/structural analyses by AFM/TEM showed that the $MoS_2$ membrane is almost conformal to the Au nanocrystals topography, with some locally detached regions. Furthermore, atomic scale resolution STEM demonstrated a variability from ~4.0 to ~4.5 Å in the Mo-Au atomic distance between 1L-$MoS_2$ and the topmost Au atoms, depending on the local configuration of Au nanocrystals planes. *Ab initio* DFT calculations of a free-standing $MoS_2$ and an ideal (unstrained) $MoS_2$/Au(111) system showed a significant influence of the Au substrate on the $MoS_2$ energy band structure, a slight improvement of the adsorption energy of the TDMAHf precursor on $MoS_2$ surface, whereas no major effects on the adsorption energy of $H_2O$ and TMA molecules. This suggests a crucial role of nanoscale morphological effects, such as the experimentally observed local curvature and strain of the $MoS_2$ membrane, in the enhanced physisorption of the precursors. Afterwards the nucleation and growth of $Al_2O_3$ an $HfO_2$ films onto 1L-$MoS_2$/Au was investigated in details, by monitoring the surface coverage as a function of the number (N) of ALD cycles, with N from 10 to 120. At low N values, $HfO_2$ exhibits a smaller coverage degree with respect to $Al_2O_3$, indicating a slower growth rate of the initially formed nuclei, probably associated to the bulky nature of the TDMAHf precursor as compared to TMA. On the other hand, the formation of continuous films was obtained in both cased for N>80 ALD cycles, corresponding to ~3.6 nm $Al_2O_3$ and ~3.1 nm $HfO_2$. Current mapping by conductive AFM on these ultra-thin films showed, under the same applied bias, a uniform insulating behavior of $Al_2O_3$ and the occurrence of few localized breakdown spots in the case of $HfO_2$, associated to a less compact films regions. Finally, an increase of the 1L-$MoS_2$ tensile strain was observed by Raman mapping after encapsulation with both high-κ films, accompanied by a reduction in the PL intensity and spectral shape, consistently with the strain and the higher effective dielectric constant of the surrounding environment.

These results demonstrate the highly beneficial role of the Au substrate for the ALD growth of uniform and ultra-thin high-k dielectrics ($Al_2O_3$ and $HfO_2$) on monolayer $MoS_2$. An optimized transfer procedure of the high-k/$MoS_2$ stack to insulating or semiconducting substrates will be the key for the extensive application of this system in electronics and optoelectronics.

ACKNOWLEDGMENTS

The authors acknowledge S. Di Franco (CNR-IMM) for assistance in the sample preparation and P. Fiorenza and G. Greco (CNR-IMM) for useful discussions.

The paper has been supported, in part, by MUR in the framework of the FlagERA- JTC 2019 project "ETMOS", by the CNR/HAS bilateral project GHOST-III (2023-25), and by European Union (NextGeneration EU), through the MUR-PNRR project SAMOTHRACE (ECS00000022).

E.S. acknowledges the PON project EleGaNTe (ARS01_01007) for financial support.

B.P and Gy. R. acknowledge funding from the national project TKP2021-NKTA-05.

A.M.M and G. N. acknowledge funding from the European Union's Horizon 2020 research and innovation programme under grant agreement No 823717 – ESTEEM3.

Part of the experiments was carried out using the facilities of the Italian Infrastructure Beyond Nano.